\documentclass[]{article}
\usepackage{amsmath}
\usepackage{amssymb}
\usepackage{amsfonts}
\usepackage{graphicx}

\begin{document}

\title{An approximate analytical solution of free convection problem for
vertical isothermal plate via transverse coordinate Taylor expansion  }
\author{Sergey Leble and Witold M.Lewandowski* \\
Gda\'{n}sk University of Technology\\
Department of Differential Equations and Applied Mathematics\\
Department of Chemical Apparatus and Machinery}
\maketitle

\begin{abstract}
The model under consideration is based on approximate analytical solution of
two dimensional stationary Navier-Stokes and Fourier-Kirchhoff equations.
Approximations are based on the typical for natural convection assumptions:
the fluid noncompressibility and Bousinesq approximation. We also assume
that ortogonal to the plate component ($x$) of velocity is neglectible
small. The solution of the boundary problem is represented as a Taylor
Series in $x$ coordinate for velocity and temperature which introduces
functions of vertical coordinate ($y$), as coefficients of the expansion.
The correspondent boundary problem formulation depends on parameters
specific for the problem: Grashoff number, the plate height ($L$) and
gravity constant. The main result of the paper is the set of equations for
the coefficient functions for example choice of expansion terms number. The
nonzero velocity at the starting point of a flow appears in such approach \
as a development of convecntional boundary layer theory formulation.
\end{abstract}

\section{Introduction}

A conventional boundary layer theory of fluid flow used for free convective
description assumes zero velocity at leading edge of a heated plate. More
advanced \ theories of self-similarity also accept this same boundary
condition \cite{1}, \cite{2}, \cite{3}. However experimental visualization
definitely shows that in the vicinity of edge the fluid motion exists \cite%
{SB}, \cite{6}, \cite{7}. It is obvious from the point of view of the mass
conservation law. In the mentioned convection descriptions the continuity
equation is not taken into account that diminishes the number of necessary
variables. For example the pressure is excluded by cross differentiation of
Navier-Stokes equation component.

The consequence of zero value of boundary layer thickness at the leading
edge of the plate yields in infinite value of heat transfer coefficient
which is in contradiction with the physical fact that the plate do not
transfer a heat at the starting point of the phenomenon. The whole picture
of the phenomenon is well known: the profiles of velocity and temperature in
normal direction to a vertical plate is reproduced by theoretical concepts
of Prandtl and self-similarity.While the evolution of profiles along tangent
coordinate do not look as given by visualisation of isotherms (see e.g. \cite%
{GDP}). It is obvious that isotherms dependance on vertical coordinate $y$
significantly differs from power low depandance $\delta\thicksim y^{1/4}$ of
boundary layer theories .

In this article we develop the model of convective heat transfer taking into
account nonzero fluid motion at the vicinity of the starting edge. Our model
is based on explicit form of solution of the basic fundamental equations
(Navier-Stokes and Fourier-Kirchhoff )\ as a power series in dependant
variables. The mass conservation law in integral form is used to formulate a
boundary condition that links initial and final edges of the fluid flow.

We consider a two-dimensional free convective fluid flow in $x,y$ plane
generated by vertical isothermal plate of height $L$ placed in udisturbed
surrounding.

The algorithm of solution construction is following. First we expand the
basic fields, velocity and temperature in power serious of horizontal
variable $x$, it substitution into the basic system gives a system of
ordinary differential equations in $y$ variable. Such system is generally
infinite therefore we should cut the expansion at some power. The form of
such cutting defines a model. The minimal number of term in the modeling is
determined by the physical conditions of velocity and temperature profiles.
From the scale analysis of the equations we neglect the horizontal (normal
to the surface of the plate) component velocity. The minimum number of
therms is chosen as three: the parabolic part guarantee a maximum of
velocity existence while the third therm account gives us change of sign of
the velocity derivative. The temperature behavior in the same order of
approximation is defined by the basic system of equations.

The first term in such expansion is linear in $x$, that account boundary
condition on the plate (isothermic one). The coefficient, noted as $C(y)$
satisfy an ordinary differential equation of the fourth order. It means that
we need four boundary condition in $y$ variable. The differential links of
other coefficients with $C$ add two constants of integrations hence a
necessity of two extra conditions. These conditions are derived from
conservation laws in integral form.

The solution of the basic system, however, need one more constant choice.
This constant characterize linear term of velocity expansion and evaluated
by means of extra boundary condition.

In the second section we present basic system in dimensional and
dimensionless forms. By means of cross-differentiation we eliminate the
pressure therm and next neglect the horizontal velocity that results in two
partial differential equations for temperature and vertical component of
velocity.

In the third section we expand both velocity and temperature fields into
Taylor series in $x$ and derive ordinary differential equations for the
coefficients by direct substitution into basic system. The minimal (cubic)
version is obtained disconnecting the infinite system of equations by the
special constraint.

The fourth and fives sections are devoted to boundary condition formulations
and its explicit form in therms of the coefficient functions of basic
fields. It is important to stress that the set of boundary conditions and
conservation laws determine all necessary parameters including the Grasshof
anf Rayleigh numbers in the stationary regime under consideration.

The last section contains the solution $C(y)$ in explicit form and results
of its numerical analysis. The solution parameters values as the function of
the plate height $L$ and parameters whivh enter the Grasshof number $%
G_{r}(l) $ estimation are given in the table form, which allows to fix a
narrow domain of the scale parameter $l$ being the characteristic linear
dimension of the flow at the starting level.

\section{The basic equations}

Let us consider a two dimensional stationary flow of incompressible fluid in
the gravity field. The flow is generated by a convective heat transfer from
solid plate to the fluid. The plate is isothermal and vertical. In the
Cartesian coordinates $x$ (horizontal and orthogonal to the palte)$,y$
(vertical and tangent to the palte) the Navier-Stokes (NS) system of
equations have the form \cite{1}.:%
\begin{equation}
\rho\left( W_{x}\frac{\partial W_{y}}{\partial x}+W_{y}\frac{\partial W_{y}}{%
\partial y}\right) =g\rho_{\infty}b\left( T-T_{\infty}\right) -\frac{%
\partial p}{\partial y}+\rho\nu\left( \frac{\partial^{2}W_{y}}{\partial y^{2}%
}+\frac{\partial^{2}W_{y}}{\partial x^{2}}\right)  \label{NSy}
\end{equation}%
\begin{equation}
\rho\left( W_{x}\frac{\partial W_{x}}{\partial x}+W_{y}\frac{\partial W_{x}}{%
\partial y}\right) =-\frac{\partial p}{\partial x}+\rho\nu\left( \frac{%
\partial^{2}W_{x}}{\partial y^{2}}+\frac{\partial^{2}W_{x}}{\partial x^{2}}%
\right)  \label{NSx}
\end{equation}

In the above equations the pressure terms are divided in two parts $%
\widetilde{p}=p_{0}+p$. The first of them is the hydrostatic one that is
equal to mass force $-g\rho_{\infty}$, where:

\begin{equation}
\rho=\rho_{\infty}\left( 1-b\left( T-T_{\infty}\right) \right)  \label{ro}
\end{equation}
is the density of  a liquid at the nondisturbed area where the temperature
is $T_{\infty}$. The second one is the extra pressure denoted by $-\nabla p.$%
The part of gravity force $gb\left( T-T_{\infty}\right) $ arises from
dependence of the extra density on temperature, $b$ is a coefficient of
thermal expansion of the fluid. In the case of gases $b=-\frac{1}{\rho}%
\left( \frac {\partial\rho}{\partial T}\right) _{p}=\frac{1}{T_{\infty}}.$%
The last terms of the above equations represents the friction forces with
the kinematic coefficient of viscosity $\nu.$

The mass continuity equation in the conditions of natural convection of
incompressible fluid in the steady state \cite{2} has the form:
\begin{equation}
\frac{\partial W_{x}}{\partial x}+\frac{\partial W_{y}}{\partial y}=0.
\label{div}
\end{equation}

The temperature dynamics is described by the stationary Fourier-Kirchhoff
(FK) equation:
\begin{equation}
W_{x}\frac{\partial T}{\partial x}+W_{y}\frac{\partial T}{\partial y}%
=a\left( \frac{\partial^{2}T}{\partial y^{2}}+\frac{\partial^{2}T}{\partial
x^{2}}\right),  \label{FK}
\end{equation}
where $W_{x}$ and $W_{y}$ are the components of the fluid velocity $%
\overline{W}$ , $T$ - temperature and $p$ - pressure disturbances
correspondingly and $a$ is the thermal diffusivity.

From the point of clarity of further transformations we use the same scale $%
l $ along both variables $x$ and $y$. We will return to the eventual
difference between characteristic scales in different directions while the
solution analysis to be provided. After introducing  variables:
\begin{equation*}
x^{\prime}=x/l,y^{\prime}=y/l,T^{\prime}=(T-T_{w})/(T_{w}-T_{\infty}),
\end{equation*}%
\begin{equation}
p^{\prime}=p/p_{\infty},W_{x}^{\prime}=W_{x}/W_{o},W_{y}^{\prime}=W_{y}/W_{o}
\label{x'}
\end{equation}
we obtain in Boussinesq approximation (in all terms besides of buoyancy one
we put $\rho\thickapprox\rho_{\infty}$ ).\bigskip%
\begin{equation}
W_{x}^{\prime}\frac{\partial W_{y}^{\prime}}{\partial x^{\prime}}%
+W_{y}^{\prime}\frac{\partial W_{y}^{\prime}}{\partial y^{\prime}}=\frac{%
gb(T_{w}-T_{\infty})l}{W_{o}^{2}}\left( T^{\prime}+1\right) -\frac{p_{\infty}%
}{\rho_{\infty}W_{o}^{2}}\frac{\partial p^{\prime}}{\partial y^{\prime}}%
+\nu^{\prime}\left( \frac{\partial^{2}W_{y}^{\prime}}{\partial y^{\prime2}}+%
\frac{\partial^{2}W_{y}^{\prime}}{\partial x^{\prime2}}\right)  \label{NS-1}
\end{equation}%
\begin{equation}
W_{x}^{\prime}\frac{\partial W_{x}^{\prime}}{\partial x^{\prime}}%
+W_{y}^{\prime}\frac{\partial W_{x}^{\prime}}{\partial y^{\prime}}=-\frac{%
p_{\infty}}{\rho_{\infty}W_{o}^{2}}\frac{\partial p^{\prime}}{\partial
x^{\prime}}+\nu^{\prime}\left( \frac{\partial^{2}W_{x}^{\prime}}{\partial
y^{^{\prime}2}}+\frac{\partial^{2}W_{x}^{\prime}}{\partial x^{\prime2}}%
\right)  \label{NS-2}
\end{equation}
and FK equation is written as
\begin{equation}
W_{x}^{\prime}\frac{\partial T^{\prime}}{\partial x^{\prime}}+W_{y}^{\prime }%
\frac{\partial T^{\prime}}{\partial y^{\prime}}=a^{\prime}\left( \frac{%
\partial^{2}T^{\prime}}{\partial y^{\prime2}}+\frac{\partial ^{2}T^{\prime}}{%
\partial x^{\prime2}}\right),  \label{FK'}
\end{equation}
where $\frac{\nu}{lW_{o}}=\nu^{\prime},\frac{a}{lW_{o}}=a^{\prime},$ $l$ is
a characteristic linear dimension and $W_{0}$ is characteristic velocity:
\begin{equation}
W_{0}=\frac{\nu}{l},  \label{W0}
\end{equation}
then $a^{\prime}=\Pr$ , $\nu^{\prime}=1$ and $\frac{gb(T_{w}-T_{\infty})l}{%
W_{0}^{2}}=G_{r},$is the Grashof number, which after plugging ($\ref{W0})$
takes the form:%
\begin{equation}
G_{r}=\frac{gb(T_{w}-T_{\infty})l^{3}}{\nu^{2}}.  \label{Gr}
\end{equation}
After cross differentiation of equations ($\ref{NS-1})$ and ($\ref{NS-2}$)
we have:
\begin{equation*}
\frac{\partial}{\partial x^{\prime}}\left[ W_{x}^{\prime}\frac{\partial
W_{y}^{\prime}}{\partial x^{\prime}}+W_{y}^{\prime}\frac{\partial
W_{y}^{\prime}}{\partial y^{\prime}}-G_{r}\left( T^{\prime}+1\right) -\left( 
\frac{\partial^{2}W_{y}^{\prime}}{\partial y^{\prime2}}+\frac{%
\partial^{2}W_{y}^{\prime}}{\partial x^{\prime2}}\right) \right] =
\end{equation*}
\begin{equation}
=\frac{\partial}{\partial y^{\prime}}\left[ W_{x}^{\prime}\frac{\partial
W_{x}^{\prime}}{\partial x^{\prime}}+W_{y}^{\prime}\frac{\partial
W_{x}^{\prime}}{\partial y^{\prime}}-\left( \frac{\partial^{2}W_{x}^{\prime}%
}{\partial y^{\prime2}}+\frac{\partial^{2}W_{x}^{\prime}}{\partial x^{\prime
2}}\right) \right]  \label{NS1+2}
\end{equation}

The FK equation rescales as 
\begin{equation}
\Pr\left( W_{x}^{\prime}\frac{\partial T^{\prime}}{\partial x^{\prime}}%
+W_{y}^{\prime}\frac{\partial T^{\prime}}{\partial y^{\prime}}\right)
=\left( \frac{\partial^{2}T^{\prime}}{\partial y^{\prime2}}+\frac {%
\partial^{2}T^{\prime}}{\partial x^{\prime2}}\right)  \label{FK''}
\end{equation}
and
\begin{equation}
\rho=\rho_{\infty}\left( 1-b\left( T-T_{\infty}\right) \right)
=\rho_{\infty}\left( 1-b\Phi\left( T^{\prime}+1\right) \right).  \label{ro1}
\end{equation}
where $\Phi=T_{w}-T_{\infty}.$

Next we would formulate the problem of free convection around the heated
vertical isothermal\ plate $x=0,$ $y\in\lbrack0,l)$, dropping the primes.
In this case we assume the angle between the plate and a stream line is
small that means a possibility to neglect the horizontal component of
velocity of fluid, denoting the vertical component as $W\left( y,x\right) $.
In this paper we restrict ourselves by the assumption that $W_{x}=0$ and $%
W_{y}=W$, that yields
\begin{equation}
\frac{\partial}{\partial x}\left[ W\frac{\partial W}{\partial y}-G_{r}\left(
T+1\right) -\left( \frac{\partial^{2}W}{\partial y^{2}}+\frac{\partial^{2}W}{%
\partial x^{2}}\right) \right] =0,  \label{NS-a}
\end{equation}
\begin{equation}
\Pr W\frac{\partial T}{\partial y}=\left( \frac{\partial^{2}T}{\partial y^{2}%
}+\frac{\partial^{2}T}{\partial x^{2}}\right) .  \label{FK-a}
\end{equation}

\section{Method of solution and approximations}

The aim of this paper is the theory application to the standard example of a
finite vertical plate. Having only two basic functions we consider the power
series expansions of the velocity and temperature in Cartesian coordinates:
\begin{equation}
W\left( x,y\right) =\gamma\left( y\right) x+\alpha\left( y\right)
x^{2}+\beta\left( y\right) x^{3}+\varkappa(y)x^{4}+....  \label{wyx}
\end{equation}
\begin{equation}
T\left( x,y\right) =C\left( y\right) x+A\left( y\right) x^{2}+B\left(
y\right) x^{3}+F\left( y\right) x^{4}+...  \label{tyx}
\end{equation}
According to standard boundary conditions on the plate we assume that the
both functions tend to zero when $x\rightarrow0$, so we choose for a
calculation the variable that has the zero value for nondimentional
temperature ($\ref{x'}$). It means that the value of $T(x,y)$ outside of the
convective flow tends to $-1.$ Substituting expressions ($\ref{wyx})$ and ($%
\ref{tyx})$ into the equations ($\ref{NS-a},\ref{FK-a}),$   we take
into account the linear independance of monomials $x^{n}$ that gives a
system of coupled nonlinear equations for the coefficients $\gamma\left(
y\right) ,$ $\alpha\left( y\right) $, ....and $C\left( y\right) $, $A\left(
y\right) $, $.....$
Such system is infinite hence for a practical use we need to choose
appropriate scheme of closed formulation for finite number of variables. The
formulation should be based on physical assumptions for a concrete
conditions.

We would like to restrict ourselves by the fourth order approximation for
both variables that means we neglect higher order terms starting from fifth
one. The area of the approximations validity is defined by the comparison of
terms in expantions ($\ref{wyx})$ and ($\ref{tyx}).$

As it will be clear from further analysis we should consider the functions: $%
\alpha\left( y\right) ,$ $\beta\left( y\right) ,$ $C\left( y\right) $ and $%
B\left( y\right) $ as variables of the first order, while $\gamma\left(
y\right) $ and $F\left( y\right) $ to be the second one. From the relations
that appear after substitution of ($\ref{wyx})$ and ($\ref{tyx})$ into ($\ref%
{NS-a})$ and ($\ref{FK-a})$ it follows that $A(y)=0$ and $F\left( y\right) =0.$
Finally from both equations ($\ref{NS-a}),$ ($\ref{FK-a})$ we obtain the
system of equations for the coefficients $B\left( y\right) $, $C\left(
y\right) $, $\alpha\left( y\right) $ and $\beta\left( y\right) :$%
\begin{equation}
6B\left( y\right) +\frac{\partial^{2}C\left( y\right) }{\partial y\partial y}%
=0,  \label{BC}
\end{equation}
\begin{equation}
\Pr\alpha\left( y\right) \frac{\partial C\left( y\right) }{\partial y}-\frac{%
\partial^{2}B\left( y\right) }{\partial y\partial y}=0,  \label{alfaBC}
\end{equation}
\begin{equation}
-6\beta\left( y\right) -G_{r}C\left( y\right) =0,  \label{betaC}
\end{equation}
\begin{equation}
\gamma\left( y\right) \frac{\partial\gamma\left( y\right) }{\partial y}-%
\frac{\partial^{2}\alpha\left( y\right) }{\partial y\partial y}=0.
\label{alfa}
\end{equation}

The first two ($\ref{BC}),$ ($\ref{alfaBC})$ arise from FK equation and the
rest of them are from the NS one. \ The system of equations is closed if $%
\gamma\left( y\right) =const=\gamma$. It means that the number of equations
and the number of unknown functions is the same.

In the first approximation the velocity and temperature are
expressed as:
\begin{equation}
W\left( y,x\right) =\gamma x+\alpha\left( y\right) x^{2}+\beta\left(
y\right) x^{3},\ \ \ \ \ T\left( y,x\right) =C\left( y\right) x+B\left(
y\right) x^{3}.  \label{twyx}
\end{equation}
From ($\ref{alfa}$) one has 
\begin{equation}
\alpha\left( y\right) =C_{1}y+C_{2}.  \label{C1C2}
\end{equation}
From ($\ref{BC}$) it follows that 
\begin{equation}
B\left( y\right) =-\frac{1}{6}\frac{\partial^{2}C\left( y\right) }{\partial
y\partial y},  \label{B}
\end{equation}
hence ($\ref{alfaBC}$) goes to:$\allowbreak$%
\begin{equation}
\frac{1}{6}\frac{\partial^{4}C\left( y\right) }{\partial y\partial y\partial
y\partial y}+\Pr\left( yC_{1}+C_{2}\right) \frac{\partial C\left( y\right) }{%
\partial y}=0.  \label{C}
\end{equation}
The equation ($\ref{betaC}$) reads:
\begin{subequations}
\begin{equation}
\beta\left( y\right) =-\frac{G_{r}}{6}C\left( y\right).  \label{BETA}
\end{equation}
This results in 
\end{subequations}
\begin{equation}
W\left( y,x\right) =\gamma x+\left( C_{1}y+C_{2}\right) x^{2}-\frac{G_{r}}{6}%
C\left( y\right) x^{3},\   T\left( y,x\right) =C\left( y\right) x-%
\frac{1}{6}\frac{\partial^{2}C\left( y\right) }{\partial y\partial y}x^{3}
\label{twyx1}
\end{equation}

The form of the equation ($\ref{C}$) indicates that for unique solution one
needs four boundary conditions for given parameters $C_{1}$ and $C_{2}$.
Apart from such conditions we should also have values for $\gamma$ and $%
G_{r} $. So for expicit determination of $W\left( y,x\right) $ and $T\left(
y,x\right) $ we need eight conditions.

\section{The analysis of the problem formulation}

Looking for the boundary conditions let us apply conservation laws of mass,
momentum and energy, applying the laws to a control volume $V$ (see Fog.1).

The first one is the conservation of mass in two dimensions that in steady
state looks as :
\begin{equation}
\int \limits_{S}\rho\overrightarrow{W}\cdot\overrightarrow{n}dS=0
\label{mass}
\end{equation}
where: $S$ is the sum of all lateral surfaces $S=\sum_{i=1}^{6}\Sigma_{i}$
(Fig.1).
\begin{figure}
	\centering
		\includegraphics{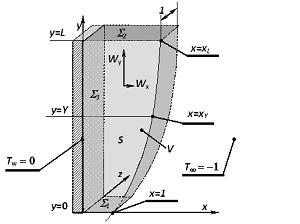}
	\caption{Fig 1. General view.}
	\label{fig:Fig-1}
\end{figure}
The mass conservation law in the integral form
($\ref{mass}$) is formulated by a division of the surface $\Sigma\ $ to two
the lower $\Sigma_{1\text{ }}$ and upper $\Sigma_{2}$ boundaries only.

According to our main assumption about two-dimensionality of the stream we
neglect a dependence of variables on $z$ coordinate and $x\in\left[ 0,1%
\right] $.
Hence\allowbreak\ the condition of total mass conservation looks as follows:
\begin{equation}
\int \limits_{\Sigma_{1}}\rho\overrightarrow{W}\cdot\overrightarrow{n}%
dS=\int \limits_{\Sigma_{2}}\rho\overrightarrow{W}\cdot\overrightarrow{n}dS
\label{totalmass}
\end{equation}
Where the flow from below $\Sigma_{1}$ is approximately the product of
density at temperature $T=-1$ and velocity of the incoming flow in the
interval $x\in\lbrack0,1].$ We follow the idea of the velocity field
continuity at $y=0$, hence $W(0,x)=\gamma x+\alpha\left( 0\right)
x^{2}+\beta\left( 0\right) x^{3}$.

For the left side in approximations mentioned above one has ($\ref{twyx1},$ $%
\ref{C1C2},$ $\ref{BETA}$ ) :

$\int \limits_{\Sigma_{1}}\rho\overrightarrow{W}\cdot\overrightarrow{n}%
dS=\rho_{\infty}\int \limits_{0}^{1}\left( \gamma x+\alpha\left( 0\right)
x^{2}+\beta\left( 0\right) x^{3}\right) dx=$\ $\rho_{\infty}\left( \frac{\gamma}{2%
}+\frac{C_{2}}{3}-\frac{G_{r}}{24}C\left( 0\right) \right) $

and outcoming flow $\Sigma_{2}$ is expressed similarily:

$\int \limits_{\Sigma_{2}}\rho\overrightarrow{W}\cdot\overrightarrow{n}%
dS=\rho_{\infty}\int \limits_{0}^{x_{L}}\left( \gamma x+\left(
C_{1}L+C_{2}\right) x^{2}-\frac{G_{r}}{6}C\left( L\right) x^{3}\right)
dx=$

$\frac{1}{24}\rho_{\infty}x_{L}^{2}\left(
12\gamma+8C_{2}x_{L}-x_{L}^{2}G_{r}C\left( L\right) +8LC_{1}x_{L}\right)
.$

The mass conservation law yields
\begin{equation}
-12\gamma-8C_{2}+8C_{2}x_{L}^{3}+G_{r}C\left( 0\right) +12\gamma
x_{L}^{2}+8LC_{1}x_{L}^{3}-x_{L}^{4}G_{r}C\left( L\right) =0.
\label{masa}
\end{equation}

The next condition is connected with the conservation of energy in
a control volume $V$ (area $S$ with unit width see Fig.1) arises from FK
equation ($\ref{FK}$) by integration over the volume.%
\begin{equation}
\Pr\int \limits_{V}\left( W\frac{\partial T}{\partial y}\right) dV=\int
\limits_{V}\left( \frac{\partial^{2}T}{\partial y^{2}}+\frac{\partial^{2}T}{%
\partial x^{2}}\right) dV=\int \limits_{S}\left(  grad T\right) 
\overrightarrow{n}dS.  \label{En}
\end{equation}

The left side of the energy conservation equation ($\ref{En}$) is
transformed similar applying the identity $ \div$($T\overrightarrow {W}%
)=T \div\overrightarrow{W}+$ $\overrightarrow{W}\cdot\nabla T$ and ($%
\ref{div}$).

According to our assumptions we left only flows accross $\Sigma_{1},\Sigma
_{2},\Sigma_{3}$  and basing on homogenity of the problem with respect to
the coordinate $z$ we have:

\begin{equation}
\int \limits_{0}^{L}\frac{\partial T}{\partial x}|_{x=0}dy+\Pr\left( -\int
\limits_{0}^{1}T(x,0)W(x,0)dx+\int \limits_{0}^{x_{L}}T(x,L)W(x,L)dx\right)
=0.  \label{En-0}
\end{equation}

To link the incoming fluid temperature $T=-1$ $\left( \text{from the bottom
edge flow}\right) $ with the solution at $y=0$ \ and the outgoing fluid (see 
$\ref{twyx1}$) we put $T(x,0)=-1$ that results in:
\begin{equation*}
\frac{C_{2}}{4}C ( L ) x_{L}^{4}-\frac{C_{2}}{36}Bx_{L}^{6}-\frac{%
 C ( L )   ^{2}}{30}  G_{r}x_{L}^{5}-\frac{ 
 x_{L}^{5}}{30}B\gamma+\frac{\gamma}{3} C ( L ) 
x_{L}^{3}-\frac{ BL}{36}C_{1}x_{L}^{6}+
\end{equation*}
\begin{equation}
\frac{ L}{4}C_{1}C ( L )
x_{L}^{4}+\frac{BG_{r}}{252}C ( L ) x_{L}^{7}+\frac{\gamma}{2}+%
\frac{C_{2}}{3}-\frac{G_{r}}{24}C\left( 0\right) +\frac{1}{\Pr}%
\int_{0}^{L}C\left( y\right) \,dy=0,  \label{EN}
\end{equation}
where $B=B(0)$. The \ equation ($\ref{C})$ is the ordinary differential
equation of the fourth order, therefore its solution needs four constants of
integration. These constants depend on two parameters $C_{1}$ and $C_{2}$,
which enter the coefficients of the Eq.($\ref{C})$. The function $C(y)$
defines the rest functions $\beta\left( y\right) $ and $B\left( y\right) $
via above relations. It means that we have six constants determining the
solution of problem and we need also six corresponding boundary conditions.

\section{Boundary conditions for temperature}

The temperature values in the vicinity of the boundary edge point and taken
as value -1 (temperature of incoming from the bottom flow). In dimensional
form the interval of consideration has the characteristic length $l$ which
we identify with a parameter we used when dimensional variables where
itroduced ($\ref{x'}$). Let us remind that scale $l$ is connected with
special (local, horizontal) Grashof number $G_{r}$ ($\ref{Gr}$).The total
height of the plate is denoted $L$

For a stationary process an edge conditions may be considered as initial one
for a Cauchy problem. Having a power series approximation of such conditions
we choose the coefficients of the series using Weierstrass theorem. It means
that we equalize the coefficients to scalar product of intial conditions and
orthonormal polynomials on the interval $\left[ 0,l\right] .$

In our case the temperature profile $T(0,x)$ represents this condition,
while the function is constant $\left( -1\right) $ on the interval $\left[
0,1\right] $ in nondimensional variables. In the approximation of the third\
power orthogonal polynomials we have:

$T\left( 0,x\right) =\allowbreak x^{3}B(0)+C(0)x=\gamma_{2}p_{1t}+\beta
_{2}p_{3t}\thickapprox-1$ because nondimentional temperature of the fluid at
the lower half plane, according to above, is $-1,$where the polynomialas are
defined as: $p_{1t}=tx,$ $\ p_{3t}=q_{t}x+u_{t}x^{3}.$

The normalization for $p_{1t}\int \limits_{0}^{1}\left( tx\right) ^{2}dx=%
\frac{1}{3}t^{2}=1$, $t=\sqrt{3},$and orthogonality condition $\int
\limits_{0}^{1}p_{1t}p_{3t}dx=\int \limits_{0}^{1}\sqrt{3}x\left(
q_{t}x+u_{t}x^{3}\right) dx=0$ give the link between constants: $\allowbreak
u_{t}=-\frac{5}{3}q_{t}$., which plugging into

$\int \limits_{0}^{1}p_{3t}^{2}dx=\int \limits_{0}^{1}\left( xq_{t}-\frac{5%
}{3}x^{3}q_{t}\right) ^{2}dx$ results in $\ q_{t}$ =$\frac{3}{2}\sqrt{7},$%
finally $p_{3t}=-\frac{x}{2}\sqrt{7}\left( 5x^{2}-3\right) .$

Substituting the result into $T\left( 0,x\right) =\allowbreak
x^{3}B(0)+C(0)x=\gamma_{2}p_{1t}+\beta_{2}p_{3t}$ gives two equations

$B\left( 0\right) +\frac{5}{2}\sqrt{7}\beta_{2}=0$, $C\left( 0\right) -\sqrt{%
3}\gamma_{2}-\frac{3}{2}\sqrt{7}\beta_{2}=0$, which solving and projecting $%
\ \beta_{2}=\int \limits_{0}^{1}x\frac{1}{2}\sqrt{7}\left( 5x^{2}-3\right)
dx=\allowbreak-\frac{1}{8}\sqrt{7}$ , $\gamma_{2}$ $=-\int \limits_{0}^{1}x%
\sqrt{3}dx=-\frac{1}{2}\sqrt{3}$.yield boundary conditions for the
coefficients for temperature expansion:
\begin{equation}
B\left( 0\right) =\frac{35}{16},C\left( 0\right) =-\frac{45}{16}.
\label{B-C}
\end{equation}
Plotting the temperature approximation at the level $y=0,$ $T\left(
0,x\right) =x^{3}B(0)+C(0)x=\allowbreak x^{3}\frac{35}{16}-\frac{45}{16}x$%
\begin{figure}
	\centering
		\includegraphics{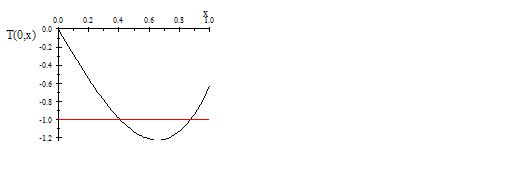}
	\caption{The fluid temperature at  $y=0$ approximation}
	\label{fig:Fig2}
\end{figure}
is given by the Fig.2.

Let us recall that $B\left( 0\right) =-\frac{1}{6}C"(0)$ (see eq. ($\ref{B}$%
))$,$ therefore $C"(0)=-\allowbreak \frac{105}{8}$ ($\ref{B-C}$) we will
consider as boundary condition for $C(y).$

The temperature gradient values $dT/dx$ on the plate decrease when $y$
grows.At the leading edge we pose the condition $\partial T/\partial
x|_{x=0}=0$ because the plate lose the contact with the fluid. It gives
third boundary condition ($\ref{twyx1}$)
\begin{equation}
C(L)=0  \label{C(L)0}
\end{equation}

\section{Boundary conditions for velocity and temperature}

The phenomenon of free convective heat transfer from isothermal vertical
plate ($T=0$) imply that temperature   gradient on the plate is negative ($%
C<0$) and decrease along $y$ ($\partial C/\partial y<0$). It is also known
that velocity profile has maximum at the distance $x_{m}$ $>0$. The extrema
for the curve is defined by derivative of $W\left( y,x\right) $ as a
function of $x.$ Hence the relation $\frac{dW}{dx}=\gamma+2\alpha x+3\beta
x^{2}=0$ indicates that for $\alpha<0$ , $\beta>0$ and $\gamma>0$ we have
two extremal points 
\begin{equation}
x_{m}=-\frac{\alpha}{3\beta}-\sqrt[2]{\frac{\alpha^{2}}{9\beta^{2}}-\frac{%
\gamma}{3\beta}}\text{ \ \ \ \ and \ \ \ \ }x_{0}(y)=-\frac{\alpha }{3\beta}+%
\sqrt[2]{\frac{\alpha^{2}}{9\beta^{2}}-\frac{\gamma}{3\beta}}  \label{xm0}
\end{equation}
if $\frac{\alpha^{2}}{9\beta}-\frac{\gamma}{3}>0.$
Notations are chosen to mark maximum position point as $x_{m}$ while the
minimum one is $x_{0}(y)>x_{m}$.

In the exeptional case of $\beta(y=L)=0,$the expression simplifies
\begin{equation}
x_{mL}=-\frac{\gamma}{2\alpha(L)},  \label{x0}
\end{equation}
which is positive for $\alpha <0.$The second extremum do not exist now (see
Fig.3).
\begin{figure}
	\centering
		\includegraphics{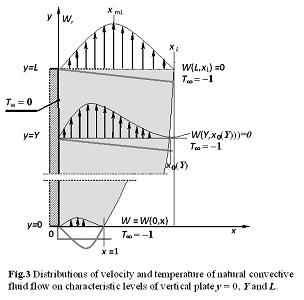}
	\label{fig:Fig-3}
\end{figure}
There is a possibility to choose the value $W(y,x_{0})=0$ considering the $%
x_{0}$ as a conditional boundary of the upward stream.We define hence $%
x_{L}=2x_{mL}$.

At the starting horizontal edge of the vertical plate the vertical velocity
component of incoming flow ($\ref{twyx1}$) varies slow so we assume that 
\begin{equation}
C_{1}=0  \label{C1}
\end{equation}
hence
\begin{equation}
W\left( y,x\right) =\gamma x+C_{2}x^{2}-\frac{G_{r}}{6}C\left( y\right)
x^{3}.  \label{Wyx}
\end{equation}
The extrema of the velocity profile ($\ref{xm0}$) after account of ($\ref{C1}
$) and ($\ref{BETA}$) is transformed as, for maximum: $x_{m}(y)=\frac{2C_{2}%
}{G_{r}C\left( y\right) }-\sqrt[2]{\frac{C_{2}{}^{2}}{\left( \frac{G_{r}}{2}%
C\left( y\right) \right) ^{2}}+\frac{2\gamma}{G_{r}C\left( y\right) }}$ and
minimum one: $x_{0}(y)=\frac{2C_{2}}{G_{r}C\left( y\right) }+\sqrt[2]{\frac{%
C_{2}{}^{2}}{\left( \frac{G_{r}}{2}C\left( y\right) \right) ^{2}}+\frac{%
2\gamma}{G_{r}C\left( y\right) }}$. The following identity $x_{0}^{2}=\frac{2%
}{G_{r}C\left( y\right) }\left( \gamma +2x_{0}C_{2}\right) $ holds for: $%
\gamma+2x_{0}C_{2}<0.$

Suppose there exists a level $y=Y$ at which 
\begin{equation}
W\left( Y,x_{0}\left( Y)\right) \right) =0  \label{Y}
\end{equation}
where $x_{0}(Y)\equiv x_{0Y}$ denotes the boundary layer thickness analog.
The equation ($\ref{Y}$) is solved with respect to $C\left( Y\right) $ that
gives: 
\begin{equation}
C\left( Y\right) =-\frac{3}{2\gamma}\frac{C_{2}^{2}}{G_{r}}  \label{C(Y)}
\end{equation}
as function of the problem parameters. Then plugging ($\ref{C(Y)}$) for the
expression for the $x_{0Y}$ yields
\begin{equation}
x_{0Y}=-2\frac{\gamma}{C_{2}}  \label{x0Y}
\end{equation}
Let us return to the expression for the temperature ($\ref{twyx1}$) with
neglecting the last term in temperature (the possibility of such assumption
will be explained below) on the level $Y$ and substitute ($\ref{C(Y)}$) and (%
$\ref{x0Y}$) into it equalizing to the temperature of surraunding $(T=-1)$ .
\begin{equation}
T\left( Y,x=x_{0Y}\right) =C\left( Y\right) x_{0Y}=-1,  \label{Tx0}
\end{equation}
we have:
\begin{equation}
C_{2}=-\frac{G_{r}}{3},\allowbreak x_{0Y}=6\frac{\gamma}{G_{r}}=6a,C\left(
Y\right) =-\frac{1}{6a}  \label{C-2}
\end{equation}
where: 
\begin{equation}
a=\frac{\gamma}{G_{r}}.  \label{a-}
\end{equation}
From the equation ($\ref{C}$) after plugging $C_{2}$ ($\ref{C-2}$) and
taking into account $C_{1}=0$ ($\ref{C1}$) we have
\begin{equation}
\frac{1}{2}\frac{\partial^{4}C\left( y\right) }{\partial y\partial y\partial
y\partial y}-\Pr G_{r}\frac{\partial C\left( y\right) }{\partial y}=0
\label{Rown}
\end{equation}
The equation was studied recently \cite{1} where the solution was given by$:$
\begin{equation}
C(y)=A_{0}+A_{1}\exp[sy]+\exp[-\frac{sy}{2}](B_{1}\cos[\frac{\sqrt{3}}{2}sy]%
+B_{2}\sin[\frac{\sqrt{3}}{2}sy]),  \label{C(y)}
\end{equation}
where 
\begin{equation}
s=\sqrt[3]{2\Pr G_{r}}  \label{s}
\end{equation}
is expressed via $Ra$ = $G_{r}\Pr.$We have also boundary conditions : ($\ref%
{B-C},\ref{C(L)0}$)

$%
\begin{array}{c}
C\left( 0\right) =\allowbreak A_{0}+A_{1}+B_{1}=-\frac{45}{16}, \\ 
C^{\prime\prime}\left( 0\right) =-6B(0)=\frac{1}{2}s^{2}\left( 2A_{1}-B_{1}-%
\sqrt{3}B_{2}\right) =-\allowbreak\frac{105}{8}, \\ 
C\left( L\right) =A_{0}+e^{-\frac{1}{2}Ls}\left( B_{1}\cos\frac{1}{2}\sqrt{3}%
Ls+B_{2}\sin\frac{1}{2}\sqrt{3}Ls\right) +\allowbreak A_{1}e^{Ls}=0.%
\end{array}
$

Solution of the system results in a rather big expression for $A_{1}\ \ $as
function of $A_{0}$which we skip in theis text, going to following
approximtion.
The explicit form of the equation ($\ref{C(y)})$ shows that the three last
terms have exponential behaviour as function of $sy.$It means that there are
three \ different domains \ of the fluid flow structure. The first is the
starting one where all terms are significant. The leading edge is
characterized by two first terms and the medium domain is described by the
only first one. We choose the parameter $y=Y$ such that it belongs to that
medium range. In such conditions
\begin{equation}
A_{0}=C\left( Y\right) =-\frac{1}{6a},  \label{A0}
\end{equation}
$\ $
where $a=\frac{\gamma}{G_{r}}.$

Plugging $A_{0}$ in the form of ($\ref{A0})$ into the table of boundary
conditions gives

$%
\begin{tabular}{l}
$\ A_{1}=-\frac{45}{16}-B_{1}+\frac{1}{6a},$ \\ 
$B_{1}=\frac{1}{9a}-\frac{1}{3}\sqrt{3}B_{2}+\frac{35}{4s^{2}}-\frac{15}{8},$
\\ 
$B_{2}=\frac{\left( e^{Ls}\left( -\frac{1}{18a}+\frac{35}{4s^{2}}+\frac {15}{%
16}\right) +\frac{1}{6a}-\left( \cos\frac{1}{2}\sqrt{3}Ls\right) e^{-\frac{1%
}{2}Ls}\left( \frac{1}{9a}+\frac{35}{4s^{2}}-\frac{15}{8}\right) \right) }{%
e^{-\frac{1}{2}Ls}\left( \sin\frac{1}{2}\sqrt{3}Ls-\frac{1}{3}\sqrt{3}\cos%
\frac{1}{2}\sqrt{3}Ls\right) +\frac{1}{3}\sqrt{3}e^{Ls}}.$%
\end{tabular}
\ \ $

Let us consider the natural approxmation $e^{-Ls}<<1$. After substitution of
the expression for $B_{2}$ into $B_{1}$and next $B_{1}$ into $\ A_{1}$ we
have approximate formulas:

$%
\begin{tabular}{l}
$A_{0}=-\frac{1}{6a},$ \\ 
$\ A_{1}=\frac{1}{6ae^{Ls}},$ \\ 
$B_{1}=\frac{1}{6a}-\frac{45}{16},$ \\ 
$B_{2}\thickapprox\allowbreak$ $\frac{35}{4}\frac{\sqrt{3}}{s^{2}}-\frac {1}{%
18}\frac{\sqrt{3}}{a}+\frac{15}{16}\sqrt{3}+\frac{1}{2a\sqrt{3}e^{Ls}}.$ $%
\allowbreak$%
\end{tabular}
$

It defines the expression for $C\left( y\right) $ as the function of
parameters $s$ ($\ref{s})$, the plate height $L$ and the new one ($\ref{a-})$
$a.$ The velocity profile at the level $Y$ is defined by ($\ref{Wyx})$ and
the parameters values ($\ref{C-2})$ :
\begin{equation}
W\left( Y,x\right) =\allowbreak\frac{1}{36}xG_{r}\frac{\left( 6a-x\right)
^{2}}{a}  \label{WYx}
\end{equation}

\section{Conservation laws application}

The mass conservation equation ($\ref{masa}$) after substitution of $%
C_{1}=0, $ $2x_{L}=3a$ $(\ref{x0}),$ $C\left( L\right) =0$, $C_{2}=-gG_r/3$ and denoting $\gamma=aG_{r}$ has the form:
\begin{equation}
\frac{1}{48}G_{r}\left( 12a-7\right) \left( 84a+144a^{2}+1\right) =0.
\label{a}
\end{equation}
The only real solution of the equation ($\ref{a}$) value that have physical
sense is $a=7/12.$

Now we can return to the energy conservation equation ($\ref{EN})$ plugging
the boundary conditions for the domain restricted by the plate on interval ($%
0,Y$). It simpifies the expression for the integral along the plate
surface (heat transfer from the plate on this interval). Consequently we
change $x_{L}$ to $x_{0Y}$ and neglect the integrant oscilations at vicinity
of $y=0$.

\begin{equation}
\frac{1}{\Pr}\int \limits_{0}^{Y}C ( y ) dy-\frac{G_{r}}{30}%
  C ( Y )  ^{2}x_{0Y}^{5}+\frac{C_{2}}{4}C\left(
Y\right) x_{0Y}^{4}+ \frac{\gamma}{3} C\left( Y\right)
x_{0Y}^{3}+\frac{\gamma}{2}+\frac{C_{2}}{3}-\frac{G_{r}}{24}C\left(
0\right) \allowbreak=0.  \label{EN'}
\end{equation}
We estimate the heat flux integral from the plate as
\begin{equation}
\frac{1}{\Pr}\int \limits_{0}^{Y}C\left( y\right) dy\thickapprox-\frac{1}{6}%
\frac{Y}{\Pr a}  \label{HF}
\end{equation}
and take into account the expressions for parameters $C_{2}=-\frac{G_{r}}{3}%
,\allowbreak x_{0Y}=6a,C\left( Y\right) =-\frac{1}{6a}$that yields:

$Ra=\frac{1}{6}\frac{Y}{a\left( \frac{1}{2}a-\frac{6}{5}a^{3}+\frac{7}{1152}%
\right) }$.

As further considerations show, the value of $Y$ mayby chosen as close to
the plate height $L$.
\begin{equation}
Ra=\frac{1}{6}\frac{L}{a\left( \frac{1}{2}a-\frac{6}{5}a^{3}+\frac{7}{1152}%
\right) }=G_{r}\cdot\Pr  \label{Ra}
\end{equation}

\section{Numerics}

Chosing $L=10$ and plugging the values of the parameters $a=7/12,\allowbreak$
$Ra=4.\,\allowbreak798\cdot L.$and ($\ref{s}$) $s=\sqrt[3]{2Ra}=\allowbreak
4.\,\allowbreak5782$ into the table of the function $C(y)$ coefficients gives

\begin{tabular}{l}
$A_{0}=-0.286,$ \\ 
$\ A_{1}=3.\,\allowbreak67\,9\times10^{-21},$ \\ 
$B_{1}=-2.\,\allowbreak527,$ \\ 
$B_{2}\thickapprox$ $2.\,\allowbreak18.$ :%
\end{tabular}

Substitution of the table values into ($\ref{C(y)}$) we have the expression
which allows to plot the function $C(y)$. 
\begin{figure}
	\centering
		\includegraphics{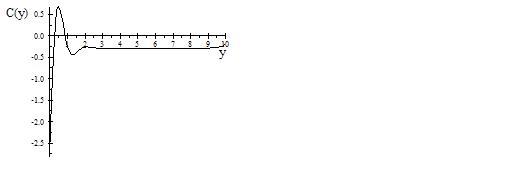}
	\caption{Fig. 4. The basic function C(y)}
	\label{fig:Fig4}
\end{figure}
In the same approximation the typical velocity profile $W\left( Y,x\right) $
($\ref{WYx}$) at the the stability interval (y $\in\left[ 2,9.5\right]: $)
$W\left( Y,x\right) =\allowbreak\frac{1}{21}xG_{r}\left( x-\frac{7}{2}%
\right) ^{2}$. Substitution of the Grashof number $Gr$= $\frac{Ra}{\Pr }=%
\frac{4.\,\allowbreak798\cdot10}{0.7}.=\allowbreak68.\,\allowbreak54$ gives
$W\left( Y,x\right) =\frac{68.\,\allowbreak 543}{21}x\left( x-\frac{7}{2}%
\right) ^{2}$ that is represented by the plot. 
\begin{figure}[h]
	\centering
		\includegraphics{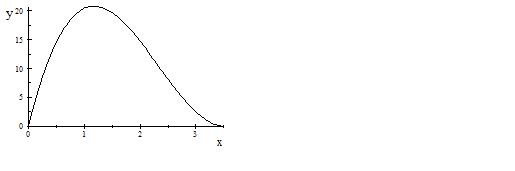}
	\caption{Fig. 5. The velocity profile at the stable region $y=Y$.}
	\label{fig:Fig6}
\end{figure}
In the same condition the temperature profile $T\left( Y,x\right) $ is
defined by the expression ($\ref{twyx1}$) and results in the plot
\begin{figure}
	\centering
		\includegraphics{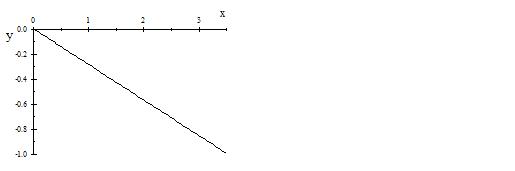}
	\caption{Fig 6. The temperature profile at $y=Y$}
	\label{fig:Fig5}
\end{figure}
To understand the phenomenon it is useful to return to dimensional picture.
As a main space scale it is choosen the parameter $l$ ($\ref{x'}$) which is
connected with the Grashof number by ($\ref{Gr}$) $l=\sqrt[3]{\frac{1}{%
bg\Phi }\nu^{2}G_{r}}$ . where $\Phi=T_{w}-T_{\infty}$. For the air example
and the temperature $T_{w}=40^{o}C,$ $T_{\infty}=20^{o}C$, $T_{av}=30^{o}C$
, $\Phi=20K$ \ the viscosity coeficient $\nu$ = $16\cdot10^{-6}m^{2}/s,$ the
coeficient of thermal expansion $b=\frac{1}{303K}$ and for conditions of our
model ($L=10$) $\allowbreak Gr=68.\,\allowbreak54$ we estimate $l$ \ as: $l=%
\sqrt[3]{\frac{1}{\frac{1}{303}10\cdot20}\left( 16\cdot10^{-6}\right)
^{2}\cdot68.\,\allowbreak54}=\allowbreak2.\,\allowbreak985\times
10^{-3}m\thickapprox3mm$

\section{Conclusions}

First of all we would stress again that the model we present here have the
engieering character of approximations, but include direct possibilities for
a development by simple taking next terms of expansions into account. A
modification of boundary conditions which would improve the transient
regimes at both ends of the y-dependence is also possible.

Newertheless in this simple modeling we observe some important
characteristic features of real convection phenomenon as\ almost parallel
streamlines and isotherms in the stability region (as, for example in \
visualizations of interferometric study from \cite{GDP} ). It follows from
functional parameter $C(y)$ behaviour inside the domain and small
contribution of cubic therm in the expresion for temperature ($\ref{twyx1}$).

Our explicit solution form and parameter values estimation allows to
conclude that:

1. the streamlines and isotherms of the flow are almost paralel to the
vertical heating plate surface in the domain of stability,

2. velocity values of the fluid flow at starting edge of the plate are
nonzero,

3. the set of boundary conditions yields in the complete set of the solution
parameter including the local Grashof number and hence, the characteristic
linear dimension length $l$ in normal to plate direction $x$,

4. the sesults allow to descibed the natural heat transfer phenomenon for
given fluid in therms only the temperature difference $\Phi$ and the plate
heigth $L,$

which are novel in comparison with former theories.


\section{References}

\begin{thebibliography}{99}
\bibitem{1} Y.Jaluria.:Natural Convection Heat and Mass Transfer; Pergamon
Press, Oxford, 1980.

\bibitem{2} Latif M. Jiji, Heat Convection, Springer-Verlag Berlin
Heidelberg, 2009.

\bibitem{3} M.Favre-Marinet, S.Tardu, Convective Heat Transfer. Solved
Problems, ISTE Ltd, John Wiley \& Sons, Inc.,2009

\bibitem{SB} E..Schmidt and W.Beckmann, Das Temperatur- und
Geschwindigkeitfeld vor einer W\={a}rme abgebenden \ senkrechten Plate bei
nat\H{u}rlicher Konvektion, Tech Mech. u. Thermodynamik, Bd.1, Nr. 10,
Okt.1930, pp.341-349 and Bd. 1, Nr.11, Nov. 1930, pp.391-406.

\bibitem{GDP} B.Gebhart, R.P.Dring, C.E.Polymeropoulos, Natural convection
from vertical surfaces, the convection transient regime, Journal of Heat
Transfer , 1967, 53-59.

\bibitem{4} Leble S., Lewandowski W.M. On analytical solution of stationary
two dimensional boundary problem of natural convection, ArXive, math, 2012.

\bibitem{5} S. Tieszen, A. Ooi, P. Durbin and M. Behnia. Modeling of natural
convection heat transfer. Center for Turbulence Research, Proceedings of the
Summer Program 1998, pp.287-302.

\bibitem{6} H.C.Li and G. P. Peterson, Experimental Studies of Natural
Convection Heat Transfer of Al2O3/DIWater Nanoparticle Suspensions
(Nanofluids), Hindawi Publishing Corporation, Advances in Mechanical
Engineering, Volume 2010, Article ID 742739

\bibitem{7} W.M. Lewandowski: Natural convection heat transfer from plates
of finite dimensions, Int. J. Heat and Mass Transfer, 34, 3, pp. 875-885,
1991.

\bibitem{YJ} Kwang-Tzu Yang and Edward W.Jerger, First-order perturbations
of laminar free-convection boundary layers on a vertical plate, Transactions
of the ASME Journal of Heat Transfer, 1964, Feb, pp.107-115.
\end{thebibliography}
\end{document}